\documentclass[a4paper]{article}
\usepackage{mathrsfs}
\usepackage{graphicx}
\usepackage{latexsym}
\usepackage{amsmath}
\usepackage{amssymb}
\usepackage{textcomp}
\usepackage{amsbsy}
\usepackage{color}
\usepackage{epstopdf}

\title{\large \bf Fermion localization in higher curvature and scalar-tensor theories of gravity}
\author{Joydip Mitra\footnote{E-mail address: jmscphys@gmail.com}\\
Department of Physics,\\ 
Scottish Church College,\\
1 $\&$ 3 Urquhart Square,
Kolkata - 700006, India.\\[10mm]
Tanmoy Paul\footnote{E-mail address: pul.tnmy9@gmail.com}\\
Soumitra SenGupta\footnote{E-mail address: tpssg@iacs.res.in}\\
Department of Theoretical Physics,\\
Indian Association for the Cultivation of Science,\\
2A $\&$ 2B Raja S.C. Mullick Road,\\
Kolkata - 700 032, India.\\[10mm]}
\date{}
\begin{document}
\maketitle

\begin{abstract}
It is well known that in a braneworld model, the localization of fermions on lower dimensional submanifold 
(say a TeV 3-brane) is governed by the gravity in the bulk which also determines the corresponding phenomenology 
on the brane. Here we consider a five dimensional warped spacetime where the bulk geometry 
is governed by higher curvature like F(R) gravity. 
In such a scenario, we explore the role of higher curvature terms 
on the localization of bulk fermions which in turn determines the effective radion-fermion coupling 
on the brane. Our result reveals that for appropriate choices of the higher curvature parameter, the profiles of the massless chiral modes 
of the fermions may get localized near TeV brane while that for massive Kaluza-Klein (KK) fermions 
localize towards the Planck brane. We also explore these features in the dual scalar-tensor model by 
appropriate transformations. The localization property turns out to be identical in both the models. 
This rules out the possibility of any signature of massive KK fermions in TeV 
scale collider experiments due to higher curvature gravity effects.
\end{abstract}
\newpage

\section{Introduction}
Over the last two decades models with extra spatial dimensions \cite{arkani,horava,RS,kaloper,cohen,burgess,chodos}  
have been increasingly playing a central role in physics beyond standard model of particle \cite{rattazzi} 
and Cosmology \cite{marteens,inflation,bouncing}. Such higher dimensional 
scenarios occur naturally in string theory. Depending on different possible compactification schemes 
for the extra dimensions, a large number of models have been constructed. In all these models, 
our visible universe is identified as one of the 3-branes embedded within a higher dimensional spacetime 
and  is described  through  a low energy effective theory on the brane carrying the signatures 
of extra dimensions \cite{kanno,shiromizu,sumanta}.\\
Among various extra dimensional models proposed over last several years, warped 
extra dimensional model pioneered by Randall and Sundrum (RS) \cite{RS} earned a 
special attention since it resolves the gauge hierarchy problem without introducing any 
intermediate scale (between Planck and TeV) in the theory. Subsequently different variants 
of warped geometry model and the issue of modulus (also known as radion) stabilization were extensively studied in 
\cite{GW,GW_radion,csaki,julien,ssg1,ssg2,tp1,tp2,tp3,sumanta_cosmology,anjan}. A generic feature of many of these models is that 
the bulk spacetime is endowed with high curvature scale $\sim$ 4 dimensional Planck scale.\\
It is well known that Einstein-Hilbert action can be generalized
by adding higher order curvature terms 
which naturally arise from diffeomorphism property of the action. 
Such terms also have their origin in String theory from  
quantum corrections. In this context $F(R)$ \cite{cruz,nojirinew,laurentis,odintsovnew,faraoni,felice,paliathanasis}, 
Gauss-Bonnet (GB) \cite{nojiri2,nojiri3,cognola}
or more generally Lanczos-Lovelock gravity 
are some of the candidates in higher curvature gravitational 
theory.\\ 
In general the higher curvature terms are suppressed with respect to Einstein-Hilbert term 
by Planck scale. Hence in low curvature regime, their contributions are negligible. However 
higher curvature terms become extremely relevant at the regime 
of large curvature. Thus for bulk geometry where the curvature is of the  
order of Planck scale, the higher curvature terms should play 
a crucial role. Motivated by this idea, in the present work, we consider 
a generalized 
warped geometry model by replacing Einstein-Hilbert bulk gravity action with a higher 
curvature $F(R)$ gravitational theory \cite{faraoni,felice,barrow,marino,bahamonde,catena,sumanta_higher,sumanta_lhc}. 
In such a scenario, several models have been studied to explore their phenomenological implications.  
Some of these are formulated by placing the standard model fields inside the bulk .
For these models the  localization property of a bulk fermion field on brane \cite{bajc,smyth,tp_fermion,chang,koley,ssg3,grossman,jm}
has been a subject of great interest to explore the chiral nature of massless 
fermion and also the hierarchial masses of fermions among different generations \cite{grossman}. 
In this context, it is observed that the overlap of the bulk fermion wave function on  
our visible brane plays crucial role in determining the effective radion-fermion coupling 
which in turn determines phenomenology of the radion with brane matter fields.\\
In view of above, it is worthwhile to address the role of higher curvature terms 
on fermion localization. We aim to address this in the present work.\\
Our paper is organized as follows : The mapping between higher curvature and scalar degrees of freedom is discussed in section II. 
The description of warped geometry in F(R) model and its solutions are described in section III and section IV. Section V addresses 
the localization 
property of bulk fermion field in the backdrop of higher curvature scenario and its consequences. The paper ends with some concluding remarks.

\section{Transformation of a F(R) theory to scalar-tensor theory}

In this section, we briefly describe how a higher curvature F(R) gravity model in five dimensional 
scenario can be recast into Einstein gravity with a scalar field. 
The F(R) action is expressed as,
\begin{equation}
 S=(1/2\kappa^2)\int d^4x dy \sqrt{G} F(R)
 \label{action0}
\end{equation}
where $x^{\mu} = (x^0, x^1, x^2, x^3)$ are usual four dimensional coordinate and 
$y$ is the extra dimensional spatial linear coordinate. 
$R$ is the five dimensional Ricci curvature and $G$ is the determinant of the metric. 
Moreover $\frac{1}{2\kappa^2}$ as taken as $2M^3$ where $M$ is the five dimensional 
Planck scale. Introducing an auxiliary field $A(x,\phi)$, 
above action (\ref{action0}) can be equivalently written as,
\begin{equation}
 S=(1/2\kappa^2)\int d^4x dy \sqrt{G} [F'(A)(R-A) + F(A)]
 \label{action00}
\end{equation}
By the variation of the auxiliary field $A(x,\phi)$, one easily obtains $A=R$. 
Plugging back this solution $A=R$ into action (\ref{action00}), 
initial action (\ref{action0}) can be reproduced. At this stage, perform a 
conformal transformation of the metric as
\begin{equation}
  G_{MN}(x,\phi) \rightarrow \tilde{G}_{MN} = \exp{(\sigma(x,\phi)}G_{MN}(x,\phi)
 \nonumber\\
\end{equation}
$M, N$ run form 0 to 5. $\sigma(x,\phi)$ is conformal factor and related to the auxiliary 
field as $\sigma = (2/3)\ln F'(A)$. Using this 
relation between $\sigma(x,\phi)$ and $A(x,\phi)$, 
one lands up with the following scalar-tensor action
\begin{eqnarray}
 S=(1/2\kappa^2)\int d^4x dy \sqrt{\tilde{G}} \bigg[\tilde{R} + 3\tilde{G}^{MN}\partial_M\sigma \partial_N\sigma 
 - (\frac{A}{F'(A)^{2/3}} - \frac{F(A)}{F'(A)^{5/3}})\bigg]
 \nonumber
\end{eqnarray}
where $\tilde{R}$ is the Ricci scalar formed by $\tilde{G}_{MN}$. $\sigma(x,\phi)$ is 
the scalar field emerged from 
higher curvature degrees of freedom. Clearly kinetic part of $\sigma(x,\phi)$ is
non canonical. In order to make the scalar field canonical, transform 
$\sigma$ $\rightarrow$ $\Phi(x,\phi) = \sqrt{3}\frac{\sigma(x,\phi)}{\kappa}$. In terms of 
$\Phi(x,\phi)$, the above action takes 
the following form,
\begin{equation}
 S=\int d^4x dy \sqrt{\tilde{G}} \bigg[\frac{\tilde{R}}{2\kappa^2} + \frac{1}{2}\tilde{G}^{MN}\partial_M\Phi 
 \partial_N\Phi - V(\Phi)\bigg]
 \nonumber\\
\end{equation}

where $V(\Phi) = \frac{1}{2\kappa^2}[\frac{A}{F'(A)^{2/3}} - \frac{F(A)}{F'(A)^{5/3}}]$ is 
the scalar field potential which depends on the 
form of $F(R)$. Thus the action of $F(R)$ gravity in five dimension can be transformed 
into the action of a scalar-tensor 
theory by a conformal transformation of the metric.

\section{Warped spacetime in F(R) model and corresponding scalar-tensor theory}

In the present paper, we consider a five dimensional AdS spacetime with two 3-brane scenario in F(R) model. The form of 
$F(R)$ is taken as $F(R) = R + \alpha R^2$ where $\alpha$ is a constant with square of the 
inverse mass dimension. Considering $\phi$ as the extra dimensional 
angular coordinate, two branes are located at $\phi = 0$ (hidden brane) and at $\phi = \pi$ (visible brane) respectively 
while the latter one is identified with the visible universe. Moreover the spacetime is $S^1/Z_2$ 
orbifolded along the coordinate $\phi$. 
The action for this model is :
\begin{eqnarray}
 S = \int d^4x dy \sqrt{G} \bigg[\frac{1}{2\kappa^2}(R + \alpha R^2) + \Lambda + V_h\delta(\phi) + V_v\delta(\phi-\pi)\bigg]
 \label{actionF(R)}
\end{eqnarray}
where $\Lambda (< 0)$ is the bulk cosmological constant and $V_h$, $V_v$ are the brane tensions on hidden, visible brane 
respectively.

This higher curvature like $F(R)$ model (in eqn.(\ref{actionF(R)})) can be transformed into scalar-tensor 
theory by using the technique discussed in the previous section. Performing a conformal transformation of the metric as 
\begin{equation}
 G_{MN}(x,\phi) \rightarrow \tilde{G}_{MN} = \exp{(\frac{1}{\sqrt{3}}\kappa\Phi(x,\phi))}G_{MN}(x,\phi)
 \label{conformal}
\end{equation}
the above action (in eqn.(\ref{actionF(R)})) can be expressed as a scalar-tensor theory with the action given by :
\begin{eqnarray}
 S&=&\int d^4x dy \sqrt{\tilde{G}} \bigg[\frac{\tilde{R}}{2\kappa^2} + \frac{1}{2}\tilde{G}^{MN}\partial_M\Phi \partial_N\Phi - V(\Phi) + \Lambda\nonumber\\ 
 &+&\exp{(-\frac{5}{2\sqrt{3}}\kappa\Phi)} V_h\delta(\phi) + \exp{(-\frac{5}{2\sqrt{3}}\kappa\Phi)} V_v\delta(\phi-\pi)\bigg]
 \label{action1ST}
\end{eqnarray}
where the quantities in tilde are reserved for ST theory. $\tilde{R}$ is the Ricci curvature formed 
by the transformed metric $\tilde{G}_{MN}$. $\Phi(x,\phi)$ is the scalar 
field corresponds to higher curvature degrees of freedom and $V(\Phi)$ is the scalar potential which 
for this specific choice form of $F(R)$ has the form,
\begin{eqnarray}
 V(\Phi)&=&\frac{1}{8\kappa^2\alpha} \exp{(-\frac{5}{2\sqrt{3}}\kappa\Phi)}\big[\exp{(\frac{3}{2\sqrt{3}}\kappa\Phi)} - 1\big]^2\nonumber\\
 &-&\Lambda \bigg[\exp{(-\frac{5}{2\sqrt{3}}\kappa\Phi)}-1\bigg]
 \label{scalar_potential}
\end{eqnarray}

One can check that the above potential (in eqn.(\ref{scalar_potential})) is stable for the 
parametric regime $\alpha > 0$. The stable value ($<\Phi>$) as well as the mass squared ($m_{\Phi}^2$) 
of the scalar field ($\Phi$) are given by the following two equations
\begin{equation}
 \exp{\bigg(\frac{3}{2\sqrt{3}}\kappa<\Phi>\bigg)} = \bigg[\sqrt{9 - 40\kappa^2\alpha\Lambda} - 2\bigg]
 \label{vev_phi}
\end{equation}
and
\begin{equation}
 m_{\Phi}^2 = \frac{1}{8\alpha} \bigg[\sqrt{9 - 40\kappa^2\alpha\Lambda}\bigg] \bigg[\sqrt{9 - 40\kappa^2\alpha\Lambda} - 2\bigg]^{-\frac{2}{3}}
 \label{mass_phi}
\end{equation}
Furthermore, the minimum value of the potential i.e. $V(<\Phi>)$ is non zero and serves as a cosmological constant. Thus 
the effective cosmological constant in scalar-tensor theory is $\Lambda_{eff} = \Lambda - V(<\Phi>)$ where $V(<\Phi>)$ 
is,
\begin{eqnarray}
 V(<\Phi>)&=&\Lambda + \bigg[\sqrt{9 - 40\kappa^2\alpha\Lambda} - 2\bigg]^{-\frac{5}{3}}\nonumber\\
 &\bigg[&-\Lambda + (1/8\kappa^2\alpha)\bigg[\sqrt{9 - 40\kappa^2\alpha\Lambda}-3\bigg]^2\bigg]
 \nonumber\\
\end{eqnarray}
Above form of $V(<\Phi>)$ with $\Lambda < 0$ clearly indicates that $\Lambda_{eff}$ is also 
negative.\\

\section{Solutions in scalar-tensor and in the corresponding F(R) theory}
Considering $\xi$ as the fluctuation 
of the scalar field over its vev, the final form of action for the scalar-tensor theory in the bulk can be written as,
\begin{eqnarray}
 S = \int d^4x dy \sqrt{\tilde{G}} [\frac{\tilde{R}}{2\kappa^2} + \frac{1}{2}\tilde{G}^{MN}\partial_M\xi 
 \partial_N\xi - (1/2)m_{\Phi}^2\xi^2 + \Lambda_{eff}]
 \label{action2ST}
\end{eqnarray}
where the terms up to quadratic order in $\xi$ are retained for $\kappa\xi < 1$. A detailed 
justification of neglecting the higher order terms as well as their possible effects 
will be discussed in the end part of this section.\\
Taking a negligible backreaction 
of the scalar field ($\xi$) on the background spacetime, the solution of metric $\tilde{G}_{MN}$ 
is exactly same as RS model i.e.
\begin{equation}
 d\tilde{s}^2 = e^{- 2 k|y|} \eta_{\mu\nu} dx^{\mu} dx^{\nu} - dy^2
 \label{grav.sol1.ST}
\end{equation}
where $k = \sqrt{\frac{-\Lambda_{eff}}{24M^3}}$. With this metric, the scalar field equation of motion 
in the bulk is following,
\begin{eqnarray}
 -\partial_y[\exp{(-4ky)}\partial_y\xi] + m_{\Phi}^2\exp{(-4ky)}\xi(y) = 0
 \label{eom.scalar.field}
\end{eqnarray}
where the scalar field $\xi$ is taken as function of extra dimensional coordinate only. Considering 
non zero value of $\xi$ on branes, the above equation (\ref{eom.scalar.field}) has the general 
solution,
\begin{equation}
 \xi(y) = e^{2ky} \big[Ae^{\nu ky} + Be^{-\nu ky}\big]
 \label{sol.scalar.field}
\end{equation}
with $\nu = \sqrt{4 + m_{\Phi}^2/k^2}$. Moreover $A$ and $B$ are obtained from the 
boundary conditions, $\xi(0)=v_h$ and $\xi(\pi r_c)=v_v$ as follows :
\begin{equation}
 A = v_v e^{-(2+\nu)kr_c\pi} - v_h e^{-2\nu kr_c\pi}
 \label{A}
\end{equation}
and
\begin{equation}
 B = v_h (1 + e^{-2\nu kr_c\pi}) - v_v e^{-(2+\nu)kr_c\pi}
 \label{B}
\end{equation}

It may be observed that  the scalar field degrees of 
freedom is related to the curvature as,
\begin{equation}
 \xi(y) = \frac{2}{\sqrt{3}\kappa}\ln[1 + 2\alpha R] - <\Phi>
 \label{scalar and curvature}
\end{equation}
Recall that $<\Phi> =\frac{2}{\sqrt{3}\kappa} \ln[\sqrt{9 - 40\kappa^2\alpha\Lambda} - 2]$.\\
From the above expression, we can relate the boundary values of the scalar field 
(i.e $\xi(0)=v_h$ and $\xi(\pi r_c)=v_v$) with the  Ricci scalar as,
\begin{equation}
 v_h = \frac{2}{\sqrt{3}\kappa} \ln\bigg[\frac{1 + 2\alpha R(0)}{\sqrt{9 - 40\kappa^2\alpha\Lambda} - 2}\bigg]
 \label{relation1}
\end{equation}
and 
\begin{equation}
 v_v = \frac{2}{\sqrt{3}\kappa} \ln\bigg[\frac{1 + 2\alpha R(\pi r_c)}{\sqrt{9 - 40\kappa^2\alpha\Lambda} - 2}\bigg]
 \label{relation2}
\end{equation}
where $R(0)$ and $R(\pi)$ are the values of the curvature on Planck and TeV brane respectively. 
Later on, we derive the expression of the bulk scalar curvature which in this 
scenario becomes dependent on the bulk coordinate $y$.
Thus the parameters that are used in the scalar-tensor theory  are actually related to the 
parameters of the original $F(R)$ theory.\\

It deserves mentioning that the solutions obtained in eqn. (\ref{grav.sol1.ST}) and eqn.(\ref{sol.scalar.field}) are based on 
the conditions that the bulk scalar potential is retained up to quadratic term (see eqn.(\ref{action2ST})) 
and the backreaction of the bulk scalar field is neglected on five dimensional spacetime. Both these 
conditions are followed from the assumption that $\kappa v_h< 1$. Relaxation of this assumption is crucial 
to check the status of higher order self interaction 
terms in the bulk scalar potential. Here we find the solutions of the field equations  
when $V(\Phi)$ is retained up to cubic term in $\Phi$. In this scenario, the five dimensional action 
in ST theory turns out to be,
\begin{eqnarray}
 S = \int d^4x dy \sqrt{\tilde{G}} \bigg[\frac{\tilde{R}}{2\kappa^2} + \frac{1}{2}\tilde{G}^{MN}\partial_M\xi 
 \partial_N\xi - (1/2)m_{\Phi}^2\xi^2 + \frac{g}{3}\xi^3 + \Lambda_{eff}\bigg]
 \label{action2ST_final}
\end{eqnarray}
where $g$ is the self cubic coupling of $\Phi(\phi)$ and can be easily determined from the form 
of $V(\Phi)$ presented in eqn.(\ref{scalar_potential}) as,
\begin{eqnarray}
 g = -\frac{\sqrt{3}\kappa}{16\alpha}\frac{[2\sqrt{9 - 40\kappa^2\alpha\Lambda} + 3]}{[\sqrt{9 - 40\kappa^2\alpha\Lambda} - 2]^{\frac{2}{3}}}
 \label{g}
\end{eqnarray}
Considering the metric ansatz as,
\begin{equation}
 d\tilde{s}^2 = e^{- 2 A(\phi)} \eta_{\mu\nu} dx^{\mu} dx^{\nu} - dy^2
 \label{grav.sol1.ST_final}
\end{equation}
the gravitational as well as the scalar field equations of motion take the following form,
\begin{eqnarray}
 4 A'^2(y) - A''(y) = -(2\kappa^2/3)(\frac{1}{2}m_{\Phi}^2\xi^2 - \frac{g}{3}\xi^3)
 \label{equation1_final}
\end{eqnarray}

\begin{eqnarray}
  A'^2(y)= \frac{\kappa^2}{12}\xi'^2 - (\kappa^2/6)(\frac{1}{2}m_{\Phi}^2\xi^2 - \frac{g}{3}\xi^3)
 \label{equation2_final}
\end{eqnarray}

\begin{eqnarray}
 \xi''(y) = 4A'\xi' + m_{\Phi}^2\xi(\phi) - g\xi^2(\phi)
 \label{equation3_final}
\end{eqnarray}
To determine the solutions of the above differential equations, we apply the iterative method 
by considering the form of metric determined in eqn. (\ref{grav.sol1.ST}) as the zeroth order solution. 
In the leading order correction of $\kappa v_h$, $\xi(\phi)$ and $A(\phi)$ turn out to be
\begin{eqnarray}
 \xi(y)&=&\bigg[Ae^{(2+\nu)ky} + Be^{(2-\nu)ky}\bigg]\ - \frac{\kappa v_h}{16\sqrt{3}\alpha k^2} \frac{[2\sqrt{9 - 40\kappa^2\alpha\Lambda} + 3]}
 {[\sqrt{9 - 40\kappa^2\alpha\Lambda} - 2]^{\frac{2}{3}}}*\nonumber\\
 &\bigg[&\frac{\exp{\big[2(2+\nu)ky-4\nu kr_c\pi\big]}}{m_{\Phi}^2r_c^2 + 8kr_c(2+\nu)-4(2+\nu)^2}\nonumber\\ 
 &+&\frac{2v_v}{v_h}\frac{\exp{\big[4ky-(2+\nu)kr_c\pi\big]}}{m_{\Phi}^2r_c^2 + 16kr_c- 16}\bigg]
 \label{sol_scalar_final}
\end{eqnarray}
and

\begin{eqnarray}
 A(y) = ky + \frac{\kappa^2v_h^2}{12}\bigg[e^{-4\nu kr_c\pi} e^{-2(2+\nu)ky} + e^{2(2-\nu)ky} \big(1+\frac{v_v}{v_h}e^{-(2+\nu)kr_c\pi}\big)\bigg]
 \label{sol_warp_final}
\end{eqnarray}
Thus due to the inclusion of the bulk scalar field backreaction, the warp factor gets modified and 
the correction term is proportional to $\kappa^2v_h^2$ which is indeed small for $\kappa v_h< 1$.\\

However from the solutions of scalar-tensor theory, one can extract the solution in F(R) model 
by inverse conformal transformation as indicated earlier (see eqn.(\ref{conformal})) :
\begin{eqnarray}
 ds^2&=&\exp{\big[-\frac{\kappa}{\sqrt{3}}(<\Phi>+\xi(y))\big]} \bigg[e^{- 2 A(y)} \eta_{\mu\nu} dx^{\mu} dx^{\nu} - dy^2\bigg]
 \label{grav.sol1.F(R)_final}
\end{eqnarray}
where $ds^2$ is the line element in $F(R)$ model and $\xi(y)$, $A(y)$ are 
given in eqn.(\ref{sol_scalar_final}) and eqn.(\ref{sol_warp_final}) respectively.\\

At this point, we need to verify whether the above solution of $G_{MN}$ (in eqn.(\ref{grav.sol1.F(R)_final})) satisfies the field equations 
of the original $F(R)$ theory. The five dimensional gravitational field equation for $F(R)$ theory is given by,
\begin{eqnarray}
 \frac{1}{2}G_{MN}F(R) - R_{MN}F'(R) - G_{MN}\Box F'(R) + \nabla_{M}\nabla_{N}F'(R) = -\frac{1}{2}\Lambda G_{MN}
 \label{field1}
\end{eqnarray}

In the present context, we take the form of $F(R)$ as $F(R) = R + \alpha R^2$ and thus the above field equation is simplified to the form :
\begin{eqnarray}
 \frac{1}{2}G_{MN}R - R_{MN} + \frac{\alpha}{2}G_{MN}R^2&-&2\alpha RR_{MN}\nonumber\\ 
 - 2\alpha G_{MN}\Box R + 2\alpha \nabla_{M}\nabla_{N}R&=&-\frac{1}{2}\Lambda G_{MN}
 \label{field2}
\end{eqnarray}

It may be shown that the solution of $G_{MN}$ in eqn.(\ref{grav.sol1.F(R)_final}) satisfies the above field equation to the leading order 
of $\kappa v_h$. It may be recalled that the equivalence of the chosen $F(R)$ model was transformed to the potential 
of the scalar-tensor model in the leading order of $\kappa v_h$. Thus it 
guarantees the validity of the solution of spacetime metric (i.e. $G_{MN}$) in the original $F(R)$ theory.\\

Using the metric solution given in eqn.(\ref{grav.sol1.F(R)_final}), one calculates the five dimensional Ricci scalar as follows:
\begin{eqnarray}
 R(y)&=&-20k^2 - \frac{20}{\sqrt{3}}e^{[-\frac{\sqrt{3}}{2}\kappa cr_cy k^2]}k^2\kappa c 
 - 5e^{[-2\frac{\sqrt{3}}{2}\kappa cr_cy k^2]}k^2\kappa^2 c^2
 \label{ricci_scalar_F(R)}
\end{eqnarray}
where $c$ is an integration constant. The presence of higher curvature gravity admits a general class of warped spacetime solution 
where the bulk curvature depends on the extra dimensional coordinate, which is in contrast to original RS situation where the 
bulk curvature is constant. Therefore, the present construction may lead to a new phenomenological scenario in the context of braneworld 
Physics which includes the effects of higher curvature terms present in the gravitational action. 
We now show how the higher curvature terms affect the localization of fermion field within the five dimensional spacetime.

\section{Fermion localization in F(R) theory}
  Consider a bulk massive fermion field propagating in a background spacetime 
  characterized by the action in eqn. (\ref{actionF(R)}). The lagrangian for the Dirac fermions is 
  given by 
  \begin{equation}
   \mathcal{L}_{Dirac} = \sqrt{-G} \bigg[\bar{\Psi}i\Gamma^{a}D_{a}\Psi - m_5\bar{\Psi}\Psi\bigg]
   \label{1}
  \end{equation}
  Using the metric solution in $F(R)$ model presented in eqn.(\ref{grav.sol1.F(R)_final}), the above 
  lagrangian can be written as,
  
  \begin{equation}
   \mathcal{L}_{Dirac} = e^{-4A(y)} e^{-\frac{5\kappa}{2\sqrt{3}}\big(<\Phi>+\xi(y)\big)} 
   \bigg[\bar{\Psi}i\Gamma^{a}D_{a}\Psi - m_5\bar{\Psi}\Psi\bigg]
   \nonumber
  \end{equation}
  where $a=(x^{\mu},y)$ are the bulk coordinates, $\Psi=\Psi(x^{\mu},y)$ is the bulk fermion field and $m_5$ is its mass. 
  $\Gamma^{a} = \bigg(e^{A(y)}e^{\frac{\kappa}{2\sqrt{3}}\big(<\Phi>+\xi(y)\big)}\gamma^{\mu}, 
  -i\gamma^{5}e^{\frac{\kappa}{2\sqrt{3}}\big(<\Phi>+\xi(y)\big)}\bigg)$ denotes the five dimensional gamma 
  matrices where $\gamma^{\mu}$ and $\gamma^{5}$ represent 4D gamma matrices in chiral 
  representation. Curved gamma matrices obey the Clifford algebra i.e. $[{\Gamma^{a},\Gamma^{b}] = 2G^{ab}}$. 
  The covariant derivative $D_{a}$ can be calculated by using the metric in eqn. (\ref{grav.sol1.F(R)_final}) 
  and is given by 
  \begin{eqnarray}
   &D_{\mu} = \partial_{\mu} +  \frac{i}{2}\gamma_{\mu}\gamma^{5}e^{\frac{\kappa}{2\sqrt{3}}\big(<\Phi>+\xi(y)\big)}
   \bigg(A'(y)+\frac{\kappa}{2\sqrt{3}}\xi'(y)\bigg)\nonumber\\
   &D_{5} = e^{\frac{\kappa}{2\sqrt{3}}\big(<\Phi>+\xi(y)\big)}\partial_{y}
   \nonumber
  \end{eqnarray}
  Using this set up, the Dirac lagrangian $\mathcal{L}_{Dirac}$ turns out to be,
  \begin{eqnarray}
   \mathcal{L}_{Dirac}&=&e^{-4A(y)}e^{-\frac{5\kappa}{2\sqrt{3}}\big(<\Phi>+\xi(y)\big)} \bar{\Psi}\bigg[ie^{A(y)}\nonumber\\
   &e&^{\frac{\kappa}{2\sqrt{3}}\big(<\Phi>+\xi(y)\big)}\gamma^{\mu}\partial_{\mu} 
   + \gamma^{5}\bigg[e^{\frac{\kappa}{2\sqrt{3}}\big(<\Phi>+\xi(y)\big)}\nonumber\\
   &\big(&\partial_{y}-2A'(y)-\frac{\kappa}{\sqrt{3}}\xi'(y)\big)\bigg] - m_5\bigg]\Psi
   \label{dirac lagrangian}
  \end{eqnarray}
  
  We decompose the five dimensional spinor via Kaluza-Klein (KK) mode expansion as 
  $\Psi(x^{\mu},y) = \sum \chi^n(x^{\mu})\zeta^n(y)$, where 
  the superscript $n$ denotes the nth KK mode.
  $\chi^n(x^{\mu})$ is the projection of $\Psi(x^{\mu},y)$ on the 3-brane and $\zeta^n(y)$ 
  is the extra dimensional component of 5D spinor. Left ($\chi_L$) and right ($\chi_R$) 
  states are constructed by $\chi^n_{L,R} = \frac{1}{2}(1 \mp \gamma^{5})\chi^n$. 
  Thus the KK mode expansion can be written in the following way : 
  \begin{equation}
   \Psi(x^{\mu},y) = \sum[\chi_L^{n}(x^{\mu})\zeta_L^{n}(y) + \chi_R^{n}(x^{\mu})\zeta_R^{n}(y)]
   \label{kk mode expansion}
  \end{equation}
  Substituting the KK mode expansion 
  of $\Psi(x^{\mu},y)$ in the Dirac field lagrangian given in eqn.(\ref{dirac lagrangian}), 
  we obtain the following equations of motion for $\zeta_{L,R}(y)$ as follows :
  \begin{eqnarray}
    &e&^{-A(y)}e^{-\frac{2\kappa}{\sqrt{3}}\big(<\Phi>+\xi(y)\big)}
   \bigg[\pm e^{\frac{\kappa}{2\sqrt{3}}\big(<\Phi>+\xi(y)\big)}\bigg(\partial_{y}-2A'(y)\nonumber\\ 
   &+&\frac{\kappa}{\sqrt{3}}\xi'(y)\bigg) + m_5\bigg]\zeta^n_{R,L}(y) = -m_n\zeta^n_{L,R}(y)
   \label{eom for wave function}
  \end{eqnarray}
  
  where $m_{n}$ is the mass of nth KK mode. The 4D fermions obey the canonical 
  equation of motion $i\gamma^{\mu}\partial_{\mu}\chi^n_{L,R} = m_{n}\chi^n_{L,R}$. 
  Moreover eqn.(\ref{eom for wave function}) is obtained provided the following normalization conditions 
  hold :
  \begin{equation}
   \int_{0}^{\pi} dy e^{-3A(y)}e^{-\frac{2\kappa}{\sqrt{3}}\big(<\Phi>+\xi(y)\big)}\zeta_{L,R}^{m}\zeta_{L,R}^{n} = \delta_{m,n}
   \label{norm1}
  \end{equation}
  \begin{equation}
   \int_{0}^{\pi} dy e^{-3A(y)}e^{-\frac{2\kappa}{\sqrt{3}}\big(<\Phi>+\xi(y)\big)}\zeta_{L}^{m}\zeta_{R}^{n} = 0
   \label{norm2}
  \end{equation}
  
  In the next two subsections, we discuss the localization scenario for massless 
  and massive KK modes respectively.
  
  \subsection{Massless KK mode}
  For massless mode, equation of motion of $\zeta_{L,R}$ takes the following form
  \begin{eqnarray}
   \bigg[\pm e^{\frac{\kappa}{2\sqrt{3}}\big(<\Phi>+\xi(y)\big)}\bigg(\partial_{y}-2A'(y) 
   - \frac{\kappa}{\sqrt{3}}\xi'(y)\bigg) + m_5\bigg]\zeta^n_{R,L}(y) = 0
   \label{eom for wave function1}
  \end{eqnarray}
  
  where $A(y)$ and $\xi(y)$ are given in eqn. (\ref{sol_warp_final}) and eqn.(\ref{sol_scalar_final}) respectively.  
  Using these forms of $A(y)$ and $\xi(y)$, we determine the solutions of $\zeta_{L,R}(y)$ as follows :
  \begin{eqnarray}
   \zeta_{L,R}(y) = \sqrt{\frac{k}{e^{(\frac{p}{k}+\frac{q}{2k})}\big[e^{2k\pi r_c}-1}\big]} e^{2ky}
   \exp{\bigg[\frac{p}{2k}e^{(2+\nu)ky} - \frac{q}{4k}e^{-2(2+\nu)ky}}\bigg]
   \label{solution1}
  \end{eqnarray}
  for $m_5 = 0$; and
  
  \begin{eqnarray}
   \zeta_{L,R}(y)&=&\sqrt{\frac{k \pm m_5}{e^{(\frac{p}{k}+\frac{q}{2k})}\big[e^{(2k\pm m_5)\pi r_c}-1}\big]} e^{(2k \pm m_5)y}*\nonumber\\ 
   &\exp&{\bigg[\frac{p}{2k}e^{(2+\nu)ky} - \frac{q}{4k}e^{-2(2+\nu)ky}}\bigg]
   \label{solution2}
  \end{eqnarray}
  for $m_5 \neq 0$.
  
  with $p$ and $q$ have following expressions :
  \begin{eqnarray}
   p = \frac{2}{\sqrt{3}}k\kappa \bigg[v_v e^{-(2+\nu)kr_c\pi} + e^{-4kr_c\pi}\frac{\kappa v_h k^{3/2}}{2\sqrt{3}}
   \frac{[2\sqrt{9 - 40\kappa^2\alpha\Lambda} + 3]}{\sqrt{9 - 40\kappa^2\alpha\Lambda}}\bigg] 
   \label{p}
  \end{eqnarray}

  and
\begin{eqnarray}
 q = \frac{2}{3} k\kappa^2 v_h^2 e^{-8k\pi r_c}
 \label{q}
 \end{eqnarray}

 where we use the explicit form of $m_{\Phi}^2$ (see eqn.(\ref{mass_phi})). 
 Moreover, the overall normalization constants in eqn. (\ref{solution1}) and eqn. (\ref{solution2}) 
  are determined by using the normalization condition presented earlier in eqn. (\ref{norm1}). 
  It may be noticed that left and right chiral modes have the same solution when $m_5=0$, 
  but the degeneracy between the two chiral modes are lifted in the presence of 
  non-zero bulk fermionic mass term.\\
  
  It is worthwhile to study how the localization scenario depends on the 
  higher curvature parameter ($\alpha$) as well as the bulk mass parameter ($m_5$).
  
  \subsubsection*{Effect of higher curvature parameter}
\begin{figure}[!h]
\begin{center}
 \centering
 \includegraphics[width=3.0in,height=2.0in]{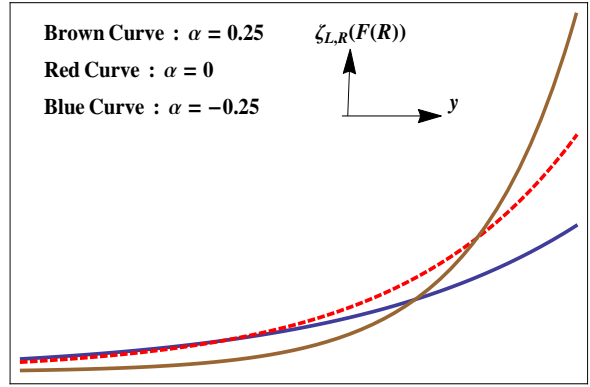}
 \caption{$\zeta_{L,R}(=\zeta_{L,R}(F(R))$ in the diagram) vs $y$ for $M=1$, $\Lambda=-1$ and $m_5=0$}
 \label{plot massless1}
\end{center}
\end{figure}

  From eqn.(\ref{solution1}) and using the expressions of $p$ and $q$, 
  we obtain Figure (\ref{plot massless1}) between $\zeta_{L,R}$ and $y$ for various values of 
the higher curvature parameter $\alpha$. We focus into the region near the TeV brane (see Figure (\ref{plot massless1})) 
to depict the localization properties of left and right chiral modes.\\

Figure (\ref{plot massless1}) clearly demonstrates that for $m_5=0$, the two chiral modes get more and more localized on TeV brane 
as the value of the parameter $\alpha$ increases. On the other hand, for small values of $\alpha$, the fermions are 
clearly localized deep inside the bulk spacetime. Thus without any bulk mass term, the fermions can be localized 
at different regions inside the bulk by adjusting the value of higher curvature parameter.\\

\begin{figure}[!h]
\begin{center}
 \centering
 \includegraphics[width=3.0in,height=2.0in]{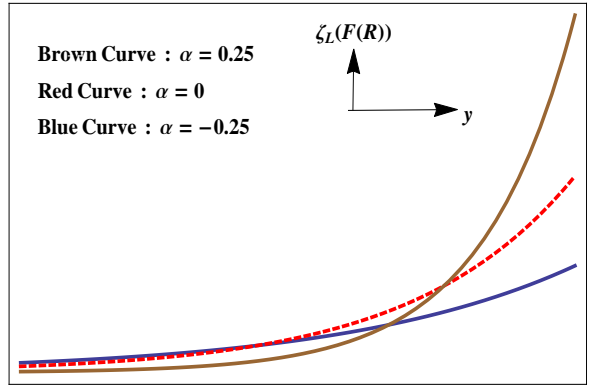}
 \caption{$\zeta_{L}(=\zeta_{L}(F(R))$ in the diagram) vs $y$ for $M=1$, $\Lambda=-1$ and $m_5=0.5$}
 \label{plot massless2}
\end{center}
\end{figure}

\begin{figure}[!h]
\begin{center}
 \centering
 \includegraphics[width=3.0in,height=2.0in]{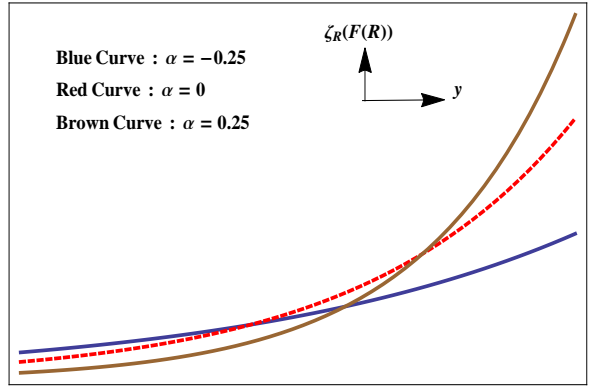}
 \caption{$\zeta_{R}(=\zeta_{R}(F(R))$ in the diagram) vs $y$ for $M=1$, $\Lambda=-1$ and $m_5=0.5$}
 \label{plot massless3}
\end{center}
\end{figure}

From eqn.(\ref{solution2}), we obtain the plots (Figure (\ref{plot massless2}) 
and Figure (\ref{plot massless3})) of 
left and right chiral massless modes 
for various values of the parameter $\alpha$ in presence of non zero bulk fermionic mass.\\

\begin{figure}[!h]
\begin{center}
 \centering
 \includegraphics[width=3.0in,height=2.0in]{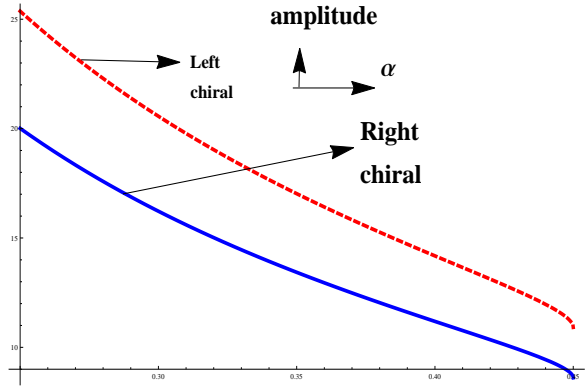}
 \caption{$\frac{\lambda_{L,R}}{m_{rad}}$ vs $\alpha$ in $F(R)$ theory for $M=1$, $\Lambda=-1$ and $m_5=0$}
 \label{plot ratio}
\end{center}
\end{figure}

Figure (\ref{plot massless2}) and Figure (\ref{plot massless3}) reveal that once again as 
the higher curvature parameter $\alpha$ increases, the peak of both left 
and right massless chiral wave function shift towards the visible brane. Thus $\alpha>0$ indicates more localization in comparison 
to RS situation. It may be mentioned that the condition $\alpha>0$ is in agreement with various astrophysical constraints 
for $F(R)=R+\alpha R^2$ as well as braneworld stability requirement [see \cite{tp1}].\\
Moreover, using the solution of $\zeta_{L,R}(y)$ (in eqn. (\ref{solution2})), 
we obtain the effective coupling \cite{rizzo} between 
radion and zeroth order fermionic KK mode as follows:
 \begin{eqnarray}
   \lambda_L&=&\sqrt{\frac{k}{24M^3}} e^{A(\pi r_c)} \bigg(\frac{e^{(k+2m_5)\pi r_c}}{e^{(k+2m_5)\pi r_c} -1}\bigg)\nonumber\\
   &\bigg[&\exp{\bigg[\frac{p}{2k}e^{(2+\nu)kr_c\pi} - \frac{q}{4k}e^{-2(2+\nu)kr_c\pi}}\bigg]\bigg]^2
   \label{coupling 4}
  \end{eqnarray}
  for left handed chiral mode and,
  \begin{eqnarray}
  \lambda_R&=&\sqrt{\frac{k}{24M^3}} e^{A(\pi r_c)} \bigg(\frac{e^{(k-2m_5)\pi r_c}}{e^{(k-2m_5)\pi r_c} -1}\bigg)\nonumber\\
   &\bigg[&\exp{\bigg[\frac{p}{2k}e^{(2+\nu)kr_c\pi} - \frac{q}{4k}e^{-2(2+\nu)kr_c\pi}}\bigg]\bigg]^2
   \label{coupling 5}
  \end{eqnarray}
  for right handed mode.\\
  With the form of $p$ and $q$ given in eqn.(\ref{p}) and eqn.(\ref{q}), it is evident that the 
  effective radion-fermion coupling increases (for both left 
and right chiral mode) with the higher curvature  
parameter $\alpha$. It is expected because the peak of both left and right chiral wave function 
get shifted towards the visible brane as $\alpha$ increases.\\
To explore the radion phenomenology, we observe that the mass of radion field in the presence of higher order curvature term ($F(R)=R+\alpha R^2$) 
  \cite{tp1}:
  \begin{eqnarray}
   m_{rad}^2 = \frac{20}{\sqrt{3}}\frac{\alpha k^4}{M^6} \epsilon^2 e^{-2kr_c\pi} v_h^2v_v^2 
   \bigg[1 + \frac{40}{\sqrt{3}}\alpha k^2\kappa v_h\bigg] \bigg[\frac{v_h}{v_v} - 1\bigg]^2
   \label{radion_mass}
  \end{eqnarray}
  It is evident that $m_{rad}$ increases with the increasing value of $\alpha$ and becomes zero as $\alpha$ tends to zero. It is expected 
  because without any higher curvature term, the gravitational action contains only Einstein-Hilbert term and thus the mass 
  of radion becomes zero (see \cite{GW_radion}).\\
  As the ratio, $\frac{\lambda_{L,R}}{m_{rad}}$ determines the fermion to radion scattering amplitude, we plot this in the figure (\ref{plot ratio}) 
  with respect to the parameter $\alpha$.\\
  
  It is evident from figure (\ref{plot ratio}) that the contribution of radion in scattering amplitude of fermions 
  decreases as the value of the higher curvature parameter $\alpha$ increases. Thus the presence of higher curvature term 
  reduces the signature in such scattering processes.\\
  
  \subsection{Massive KK mode}
In this section, we study the localization of higher Kaluza-Klein modes. For massive KK 
modes, equation of motion for fermionic wave function is given as,
\begin{eqnarray}
   &e&^{-A(y)}e^{-\frac{2\kappa}{\sqrt{3}}\big(<\Phi>+\xi(y)\big)}
   \bigg[\pm e^{\frac{\kappa}{2\sqrt{3}}\big(<\Phi>+\xi(y)\big)}\bigg(\partial_{y}-2A'(y)\nonumber\\ 
   &-&\frac{\kappa}{\sqrt{3}}\xi'(y)\bigg) + m_5\bigg]\zeta^n_{R,L}(y) = -m_n\zeta^n_{L,R}(y)
   \label{eom for wave function2}
  \end{eqnarray}
  
  $m_n$ is the mass of nth KK mode. Using the rescaling 
  \begin{eqnarray}
   \omega^{(n)}_{L,R} = e^{\frac{5}{2}A(y)} e^{\frac{5\kappa}{4\sqrt{3}}\big(<\Phi>+\xi(y)\big)}\zeta^{(n)}_{L,R}
   \label{rescale}
  \end{eqnarray}
  
  we find that the two helicity states, $\zeta_L$ and $\zeta_R$ satisfy the same equation of motion and is given by,
  \begin{eqnarray}
   &\frac{\partial^2\omega^{(n)}}{\partial y^2}& + e^{-\frac{\kappa}{\sqrt{3}}\big(<\Phi>+\xi(y)\big)}\bigg[e^{\frac{\kappa}
   {\sqrt{3}}\big(<\Phi>+\xi(y)\big)} \bigg(-\frac{1}{4}A'^2 - \frac{\kappa^2}{48}\xi'^2\nonumber\\ 
   &-&\frac{\kappa}{4\sqrt{3}}A'\xi' + m_{n}^2e^{2A(y)}\bigg) + m_5^2\bigg] \omega_{L,R} = 0
   \label{reduced eom}
  \end{eqnarray}
  
  Using the forms of $A(y)$ and $\xi(y)$, we obtain the solution of eqn.(\ref{reduced eom}) and is given by Hypergeometric 
  function as follows :
  \begin{eqnarray}
   &\zeta&^{(n)}_{L,R}(y) = \sqrt{k} \exp{\bigg[-\frac{p}{8k}e^{(2+\nu)ky}\bigg]}*\nonumber\\
   &\exp&{\bigg[-\frac{1}{8k}\sqrt{q^2e^{4(2+\nu)ky}+k^2+4m_{n}^2e^{2ky}+4m_5^2}\bigg]}*
   HypergeometricU\bigg[\frac{5}{8}\nonumber\\
   &+&\frac{qe^{2(2+\nu)ky}}{8k} + \frac{\sqrt{q^2e^{4(2+\nu)ky}+k^2+4m_{n}^2e^{2ky}+4m_5^2}}{8k},\nonumber\\
   &1&+ \frac{\sqrt{q^2e^{4(2+\nu)ky}+k^2+4m_{n}^2e^{2ky}+4m_5^2}}{4k},\frac{p}{4k}e^{(2+\nu)ky}\bigg]
   \label{solution3}
  \end{eqnarray}
  
  The mass spectrum can be obtained from the requirement that the wave function is 
  well behaved on the brane. Demanding the continuity of $\xi_{L,R}$ at $y=0$ and at $y=\pi r_c$, the KK mass term 
  can be obtained as follows :
   \begin{eqnarray}
    m_n^2&=&e^{-2A(\pi r_c)} e^{-\frac{\kappa}{\sqrt{3}}\big(<\Phi>+\xi(\pi r_c)\big)} \bigg[k^2(n^2 + 2n + 1) + m_5^2\bigg]
    \label{mass spectrum}
   \end{eqnarray}
   where $n=1,2,3....$. The above expression of mass spectrum is in agreement with \cite{jm}.\\ 
   Now from the requirement of resolving the gauge hierarchy 
   problem, the warp factor at TeV brane acquires the value $= 36$ 
   which produces a large suppression in the right hand side of eqn. (\ref{mass spectrum}) 
   through the exponential factor. Since $k, m_5\sim M$, the mass of KK modes ($n=1,2,3..$) turn out to be at the TeV scale.\\
   Using the solution of $\zeta^n_{L,R}(y)$ (in eqn. (\ref{solution3})), we determine the coupling between 
   massive KK fermion modes and the radion field as,
   \begin{eqnarray}
    &\lambda&^{(n)} = \sqrt{\frac{k}{24M^3}}e^{A(\pi r_c)} e^{\frac{\kappa}{2\sqrt{3}}\big(<\Phi>+\xi(\pi r_c)\big)}
    \exp{\bigg[-\frac{p}{4k}e^{(2+\nu)k\pi r_c}\bigg]}\nonumber\\
   &\exp&{\big[-\frac{1}{4k}\sqrt{q^2e^{4(2+\nu)k\pi r_c}+k^2+4m_{n}^2e^{2k\pi r_c}+4m_5^2}\big]}*
   \bigg[HypergeometricU\bigg[\frac{5}{8}\nonumber\\
   &+&\frac{qe^{2(2+\nu)k\pi r_c}}{8k} + \frac{\sqrt{q^2e^{4(2+\nu)k\pi r_c}+k^2+4m_{n}^2e^{2k\pi r_c}+4m_5^2}}{8k},\nonumber\\
   &1&+ \frac{\sqrt{q^2e^{4(2+\nu)k\pi r_c}+k^2+4m_{n}^2e^{2k\pi r_c}+4m_5^2}}{4k},\frac{p}{4k}e^{(2+\nu)k\pi r_c}\bigg]\bigg]^2
  \label{coupling}
 \end{eqnarray}
   where $\lambda^{(n)}$ is the coupling between $n$ th KK fermion mode and the radion field and $p$, $q$ are given in eqn.(\ref{p}), 
   eqn.(\ref{q}) respectively. With increasing $\alpha$, the third argument of Hypergeometric function increases 
   in the above expression and as a result $\lambda^{(n)}$ decreases.\\
   
\begin{figure}[!h]
\begin{center}
 \centering
 \includegraphics[width=3.0in,height=2.0in]{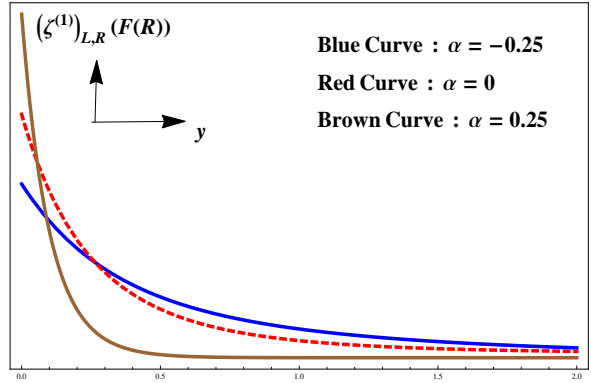}
 \caption{$\zeta^{(1)}_{L,R}(=\zeta^{(1)}_{L,R}(F(R))$ in the diagram) vs $y$ for $M=1$, $\Lambda=-1$ and $m_5=0.5$}
 \label{plot massive mode}
\end{center}
\end{figure}

   Eqn.(\ref{solution3}) indicates the relation between $\zeta_{L,R}$ and $y$ for various values of $l$ 
   from which one can find the dependence of localization 
   for massive KK fermion modes on the backreaction parameter. 
   From this, the behaviour of the first KK mode ($n=1$) is described in figure (\ref{plot massive mode}).
   
Figure (\ref{plot massive mode}) clearly depicts that the wave function for 
first massive KK mode gets more and more localized near Planck brane with increasing 
value of higher curvature parameter. Consequently, the coupling parameter decreases near the visible 
brane as $\alpha$ increases.This may explain the invisibility of massive KK mode fermions in the search for 
signature of warped extra dimensions in collider Physics.\\  
Moreover, it can also be shown (from eqn.(\ref{solution3})) that as the order of KK mode increases from $n=1$, 
the localization of fermions becomes sharper near Planck brane.\\

It may be mentioned that the bulk fermion mass term ($m_5$) also affects the localization of fermion field. 
Using the solution of $\zeta_{L,R}(y)$ presented in eqn. (\ref{solution2}), it can be shown that for a fixed value of backreaction 
parameter, the left chiral mode of zeroth KK fermion has higher peak values on TeV brane as 
the bulk fermion mass increases where as the right chiral mode 
shows the reverse nature, which is in agreement with \cite{jm,tp_fermion}.\\

\section{Fermion localization in the corresponding scalar-tensor (ST) theory}
Till now we have described the fermion localization in the original $F(R)$ gravity model. To bring out the 
equivalence in the ST theory, we need to transform the original Dirac 
lagrangian (shown in eqn.(\ref{1})) into scalar-tensor version of the model and a conformal transformation 
of $G_{MN}$ (as mentioned in eqn.(\ref{conformal})) fulfills the purpose. With such conformal transformation, 
$L_{Dirac}$ takes the following form:
\begin{eqnarray}
 L_{Dirac} = \sqrt{\tilde{G}} [e^{-\frac{2}{\sqrt{3}}\kappa\Phi} \bar{\Psi}i\tilde{\Gamma}^{a}\tilde{D}_{a}\Psi 
 - m_5e^{-\frac{5}{\sqrt{3}}\kappa\Phi} \bar{\Psi}\Psi]
 \label{1S}
\end{eqnarray}

Recall that $\tilde{G}_{MN}$ (the quantities in tilde are reserved for ST theory) is the spacetime metric in ST theory 
and given by (see eqn.(\ref{grav.sol1.ST_final})):
\begin{eqnarray}
 d\tilde{s}^2 = e^{- 2 A(\phi)} \eta_{\mu\nu} dx^{\mu} dx^{\nu} - dy^2
 \nonumber
\end{eqnarray}

where $A(y)$ is obtained in eqn.(\ref{sol_warp_final}). 
  $\tilde{\Gamma}^{a} = (e^{A(y)}\tilde{\gamma}^{\mu}, -i\tilde{\gamma}^{5})$ is the five dimensional gamma 
  matrices where $\tilde{\gamma}^{\mu}$ and $\tilde{\gamma}^{5}$ represent 4D gamma matrices in ST theory. 
  By using the form of $\tilde{G}_{MN}$, the covariant derivative $D_{a}$ can be calculated and is given by, 
  \begin{eqnarray}
   &\tilde{D}_{\mu} = \partial_{\mu} - \frac{1}{2}\tilde{\Gamma}_{\mu}\tilde{\Gamma}^{5}A'(y)e^{-A(y)}\nonumber\\
   &\tilde{D}_{5} = \partial_{y}
   \nonumber
  \end{eqnarray}
  
  Further from eqn.(\ref{1S}), it is evident that $\Psi(x^{\mu},y)$ is not canonical and thus the canonical Dirac 
  field ($\tilde{\Psi}$) in ST theory is defined as,
  \begin{eqnarray}
   \tilde{\Psi}(x^{\mu},y) = e^{-\frac{1}{\sqrt{3}}\kappa\Phi} \Psi(x^{\mu,y})
   \nonumber
  \end{eqnarray}
  
  In terms of such canonical field, the Dirac lagrangian turns out to be,
  \begin{eqnarray}
   L_{Dirac} = \sqrt{\tilde{G}} \bigg[\bar{\tilde{\Psi}}i\tilde{\Gamma}^{a}\tilde{D}_{a}\tilde{\Psi} 
 - m_5e^{-\frac{2}{\sqrt{3}}\kappa\Phi} \bar{\tilde{\Psi}}\tilde{\Psi} + \frac{1}{\sqrt{3}}\kappa\Phi'(y)\big(\bar{\tilde{\Psi}}\tilde{\gamma}^5\tilde{\Psi}\big)\bigg]
 \label{2S}
  \end{eqnarray}
  
  As earlier, we consider $\xi(y)$ as the fluctuation of the scalar field ($\Phi(y)$) over its vev ($<\Phi>$) 
  i.e $\Phi(y) = <\Phi> + \xi(y)$. Therefore $L_{Dirac}$ can be expressed as,
  \begin{eqnarray}
   L_{Dirac}&=&\sqrt{\tilde{G}} \bigg[\bar{\tilde{\Psi}}i\tilde{\Gamma}^{a}\tilde{D}_{a}\tilde{\Psi} 
 - m_5e^{-\frac{2}{\sqrt{3}}\kappa<\Phi>} \bar{\tilde{\Psi}}\tilde{\Psi}\nonumber\\
 &+&\frac{1}{2\sqrt{3}}\kappa m_5 \bar{\tilde{\Psi}}\tilde{\Psi} \xi(y) - \frac{1}{24}\kappa^2 m_5 \bar{\tilde{\Psi}}\tilde{\Psi} \xi(y)^2 
 + \frac{1}{\sqrt{3}}\kappa\xi'(y)\big(\bar{\tilde{\Psi}}\tilde{\gamma}^5\tilde{\Psi}\big)\bigg]
 \label{3S}
  \end{eqnarray}
  
  where the higher order terms of $\kappa\xi$ are neglected due to the consideration of $\kappa\xi < 1$. It may be noted that 
  the effective mass of the bulk fermion field in ST theory gets suppressed by a factor of $e^{-\frac{\kappa}{2\sqrt{3}}<\Phi>}$ 
  in comparison to that of $F(R)$ theory i.e. $\tilde{m_5} = m_5e^{-\frac{\kappa}{2\sqrt{3}}<\Phi>}$, where $\tilde{m_5}$ 
  symbolizes the bulk mass in ST theory. Moreover the last three terms in the expression of eqn.(\ref{3S}) denote the 
  coupling between scalar field ($\xi(y)$, generated from higher curvature degrees of freedom) and fermion field. 
  Clearly such couplings carry the signature of higher curvature effects.\\
  
  With the explicit forms of $\tilde{D}_{\mu}$ and $\tilde{D}_5$, $L_{Dirac}$ is finally written as,
  \begin{eqnarray}
   L_{Dirac}&=&e^{-4A(y)} \bar{\tilde{\Psi}}\bigg[ie^{A(y)}\tilde{\gamma}^{\mu}\partial_{\mu} 
   + \tilde{\gamma}^{5}\bigg(\partial_{y}-2A'(y)+\frac{\kappa}{\sqrt{3}}\xi'(y)\bigg)\nonumber\\ 
   &-&m_5\bigg(e^{-\frac{\kappa}{2\sqrt{3}}<\Phi>} - \frac{\kappa}{2\sqrt{3}}\xi(y) + \frac{\kappa^2}{24}\xi^2(y)\bigg)\bigg]\Psi
   \label{4S}
  \end{eqnarray}
  
  At this stage, the five dimensional spinor $\tilde{\Psi}(x^{\mu},y)$ is decomposed via KK mode expansion as follows,
  \begin{equation}
   \tilde{\Psi}(x^{\mu},y) = \sum[\tilde{\chi}_L^{n}(x^{\mu})\tilde{\zeta}_L^{n}(y) + \tilde{\chi}_R^{n}(x^{\mu})\tilde{\zeta}_R^{n}(y)]
   \label{5S}
  \end{equation}
  
  $n$ denotes the nth KK mode and the subscripts $L,R$ indicate the left, right chiral (or helicity) states respectively. 
  The above KK mode expansion along with the Dirac field lagrangian given in eqn.(\ref{4S}) lead the following equation of motion 
  for $\tilde{\zeta}^{(n)}_{L,R}(y)$ :
  \begin{eqnarray}
   e^{-A(y)}&\bigg[&\pm\bigg(\partial_{y} - 2A'(y) + \frac{\kappa}{\sqrt{3}}\xi'(y)\bigg)\nonumber\\ 
   &+&m_5\bigg(e^{-\frac{\kappa}{2\sqrt{3}}<\Phi>} - \frac{\kappa}{2\sqrt{3}}\xi(y) + \frac{\kappa^2}{24}\xi^2(y)\bigg)\bigg]\tilde{\zeta}^n_{R,L}(y)\nonumber\\ 
   &=&-\tilde{m}_n\tilde{\zeta}^n_{L,R}(y)
   \label{6S}
  \end{eqnarray}
  
  where $\tilde{m}_{n}$ is the mass of nth KK mode in St theory. The 4D fermions obey the canonical 
  equation of motion $i\tilde{\gamma}^{\mu}\partial_{\mu}\tilde{\chi}^n_{L,R} = \tilde{m}_{n}\tilde{\chi}^n_{L,R}$. 
  Moreover eqn.(\ref{6S}) is obtained provided the following normalization conditions 
  hold :
  \begin{equation}
   \int_{0}^{\pi} dy e^{-3A(y)}\tilde{\zeta}_{L,R}^{m}\tilde{\zeta}_{L,R}^{n} = \delta_{m,n}
   \label{norm1S}
  \end{equation}
  \begin{equation}
   \int_{0}^{\pi} dy e^{-3A(y)}\tilde{\zeta}_{L}^{m}\tilde{\zeta}_{R}^{n} = 0
   \label{norm2S}
  \end{equation}
  
  In the next two subsections, we discuss the localization scenario for massless 
  and massive KK modes (for ST theory) respectively.
  
  \subsection{Massless KK mode}
  For massless mode, equation of motion of $\tilde{\zeta}_{L,R}$ takes the following form
  \begin{eqnarray}
   e^{-A(y)}&\bigg[&\pm\bigg(\partial_{y} - 2A'(y) + \frac{\kappa}{\sqrt{3}}\xi'(y)\bigg)\nonumber\\ 
   &+&m_5\bigg(e^{-\frac{\kappa}{2\sqrt{3}}<\Phi>} - \frac{\kappa}{2\sqrt{3}}\xi(y) + \frac{\kappa^2}{24}\xi^2(y)\bigg)\bigg]\tilde{\zeta}^n_{R,L}(y) 
   = 0
   \label{7S}
  \end{eqnarray}
  
  Using the explicit solution of $\xi(y)$ and $A(y)$ (see eqn.(\ref{sol_scalar_final}) and eqn.(\ref{sol_warp_final})), we 
  obtain the solution of eqn.(\ref{7S}) as follows :
  \begin{eqnarray}
   \tilde{\zeta}_{L,R}(y) = \sqrt{\frac{k}{\big[e^{2k\pi r_c}-1\big]}} e^{2ky}
   \exp{\bigg[\frac{p}{2k}e^{(2+\nu)ky} - \frac{q}{4k}e^{-2(2+\nu)ky}}\bigg]
   \label{8S}
  \end{eqnarray}
  for $m_5 = 0$; and
  
  \begin{eqnarray}
   \tilde{\zeta}_{L,R}(y)&=&\sqrt{\frac{k \pm \tilde{m}_5}{e^{(\frac{p}{k}+\frac{q}{2k})}\big[\big(1\mp \frac{\kappa v_h}{2\sqrt{3}}
   m_5 e^{(2+\nu)k\pi r_c}\big)e^{(2k\pm m_5)\pi r_c}-1\big]}}\nonumber\\ 
   &\exp&{\bigg[(2k \pm \tilde{m}_5)y \mp \frac{\kappa v_h}{2\sqrt{3}}m_5 e^{(2+\nu)ky} \pm \frac{\kappa^2 v_h^2}{24}m_5 e^{2(2+\nu)ky}\bigg]}*\nonumber\\ 
   &\exp&{\bigg[\frac{p}{2k}e^{(2+\nu)ky} - \frac{q}{4k}e^{-2(2+\nu)ky}}\bigg]
   \label{9S}
  \end{eqnarray}
  for $m_5 \neq 0$.
  
  with $p$ and $q$ are given in eqn.(\ref{p}) and eqn.(\ref{q}) respectively. 
  Moreover, the overall normalization constants in eqn. (\ref{8S}) and eqn. (\ref{9S}) 
  are determined by using the normalization condition presented earlier in eqn. (\ref{norm1S}).\\
  
\begin{figure}[!h]
\begin{center}
 \centering
 \includegraphics[width=3.0in,height=2.0in]{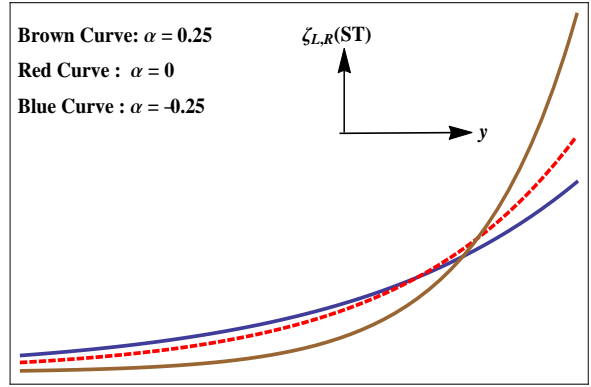}
 \caption{$\tilde{\zeta}_{L,R}(=\zeta_{L,R}(ST)$ in the diagram) vs $y$ for $M = 1$, $\Lambda = -1$ and $m_5=0$}
 \label{plot massless1S}
\end{center}
\end{figure}

  From eqn. (\ref{8S}), we obtain the Figure (\ref{plot massless1S}) between $\tilde{\zeta}_{L,R}$ and $y$ for various 
  values of $\alpha$.\\
  
The constant $y$ hypersurfaces at $y=0$ and $y=36$ represent Planck and TeV branes respectively. 
We focus into the region near the TeV brane (see figure (\ref{plot massless1S})) 
to depict the localization properties of the left and right modes.

Figure (\ref{plot massless1S}) clearly reveals that similar to $F(R)$ theory, the massless left and right chiral modes 
(for $m_5 = 0$) get more localized on TeV brane as the value of $\alpha$ increases.\\

From eqn. (\ref{9S}), we obtain the plots of left and right chiral modes for various $\alpha$ 
in presence of non-zero bulk fermionic mass.

\begin{figure}[!h]
\begin{center}
 \centering
 \includegraphics[width=3.0in,height=2.0in]{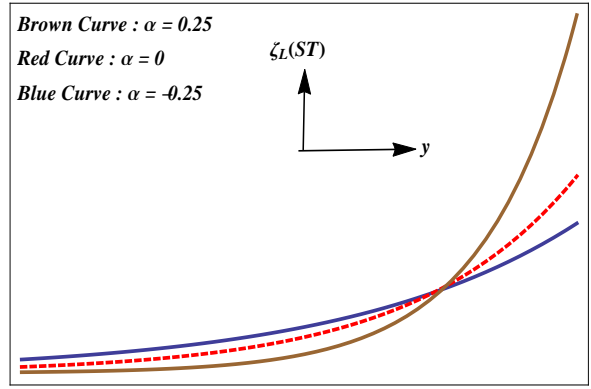}
 \caption{$\tilde{\zeta}_{L}(=\zeta_{L}(ST)$ in the diagram) vs $y$ for $M = 1$, $\Lambda = -1$ and $m_5=0.5k$}
 \label{plot massless2S}
\end{center}
\end{figure}

\begin{figure}[!h]
\begin{center}
 \centering
 \includegraphics[width=3.0in,height=2.0in]{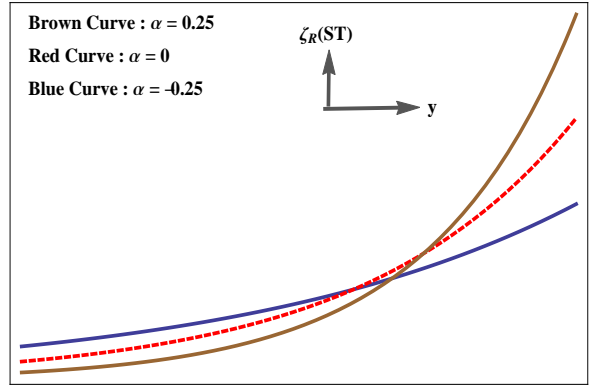}
 \caption{$\tilde{\zeta}_{R}(=\zeta_{R}(ST)$ in the diagram) vs $y$ for $M = 1$, $\Lambda = -1$ and $m_5=0.5k$}
 \label{plot massless3S}
\end{center}
\end{figure}

Figure (\ref{plot massless2S}) and figure (\ref{plot massless3S}) 
reveal that as the value of $\alpha$ increases, the peak of both left and right chiral wave 
function get shifted towards the visible brane.\\

Moreover, using the solution of $\tilde{\zeta}_{L,R}(y)$ (in eqn. (\ref{9S})), we obtain the effective coupling \cite{rizzo} between 
radion and zeroth order fermionic KK mode (in ST theory) as follows:
 \begin{eqnarray}
   \tilde{\lambda}_L&=\sqrt{\frac{k}{24M^3}} e^{A(\pi r_c)} 
   \bigg[\exp{\bigg[\frac{p}{2k}e^{(2+\nu)kr_c\pi} - \frac{q}{4k}e^{-2(2+\nu)kr_c\pi}}\bigg]\bigg]^2\nonumber\\
   &\frac{\exp{\bigg[(k + 2\tilde{m}_5)\pi r_c - \frac{\kappa v_h}
   {\sqrt{3}}m_5 e^{(2+\nu)k\pi r_c} + \frac{\kappa^2 v_h^2}{12}m_5 e^{2(2+\nu)k\pi r_c}\bigg]}}
   {\big[\big(1 - \frac{\kappa v_h}{2\sqrt{3}} m_5 e^{(2+\nu)k\pi r_c}\big)e^{(k + 2\tilde{m}_5)\pi r_c}-1\big]}
   \label{10S}
  \end{eqnarray}
  for left handed chiral mode and,
  \begin{eqnarray}
  \tilde{\lambda}_R&=\sqrt{\frac{k}{24M^3}} e^{A(\pi r_c)} 
   \bigg[\exp{\bigg[\frac{p}{2k}e^{(2+\nu)kr_c\pi} - \frac{q}{4k}e^{-2(2+\nu)kr_c\pi}}\bigg]\bigg]^2\nonumber\\
   &\frac{\exp{\bigg[(k - 2\tilde{m}_5)\pi r_c + \frac{\kappa v_h}
   {\sqrt{3}}m_5 e^{(2+\nu)k\pi r_c} + \frac{\kappa^2 v_h^2}{12}m_5 e^{2(2+\nu)k\pi r_c}\bigg]}}
   {\big[\big(1 + \frac{\kappa v_h}{2\sqrt{3}} m_5 e^{(2+\nu)k\pi r_c}\big)e^{(k - 2\tilde{m}_5)\pi r_c}-1\big]}
   \label{11S}
  \end{eqnarray}
  for right handed mode.\\
  
  Comparing the above two expressions with eqn.(\ref{coupling 4}), eqn.(\ref{coupling 5}), it is clear that the 
  coupling between radion and zeroth KK fermionic mode is different with respect to that of $F(R)$ theory 
  and the difference is indeed small for $\kappa v_h < 1$. However the effective radion-fermion coupling in ST theory  
  increases (for both left and right chiral mode) with the parameter $\alpha$, which is in agreement with the 
  corresponding $F(R)$ theory.\\





\subsection{Massive KK mode}
For massive KK modes, equation of motion for fermionic wave function is given as,
\begin{eqnarray}
   e^{-A(y)}&\bigg[&\pm\bigg(\partial_{y} - 2A'(y) + \frac{\kappa}{\sqrt{3}}\xi'(y)\bigg)\nonumber\\ 
   &+&m_5\bigg(e^{-\frac{\kappa}{2\sqrt{3}}<\Phi>} - \frac{\kappa}{2\sqrt{3}}\xi(y) + \frac{\kappa^2}{24}\xi^2(y)\bigg)\bigg]\tilde{\zeta}^n_{R,L}(y)\nonumber\\ 
   &=&-\tilde{m}_n\tilde{\zeta}^n_{L,R}(y)
   \label{12S}
  \end{eqnarray}
  
  Recall that $\tilde{m}_n$ is the mass of nth KK mode in ST theory. With the help of rescaling wave function 
  $\tilde{\omega}_{L,R} = e^{\frac{5}{2}A(y)}\tilde{\zeta}_{L,R}$, we obtain the solution of $\tilde{\zeta}^{(n)}_{L,R}$ as follows :
  \begin{eqnarray}
   &\tilde{\zeta}&^{(n)}_{L,R}(y) = \sqrt{k} \exp{\bigg[-\frac{p}{8k}e^{(2+\nu)ky}\bigg]}*\nonumber\\
   &\exp&{\bigg[-\frac{1}{8k}\sqrt{q^2e^{4(2+\nu)ky}+k^2+4\tilde{m}_{n}^2e^{2ky}+4\tilde{m}_5^2+\frac{\kappa^2 v_h^2}{3}m_5^2e^{2(2+\nu)ky}}\bigg]}\nonumber\\
   &\bigg(&HypergeometricU\bigg[\frac{5}{8} + \frac{qe^{2(2+\nu)ky}}{8k}\nonumber\\ 
   &+&\frac{\sqrt{q^2e^{4(2+\nu)ky}+k^2+4\tilde{m}_{n}^2e^{2ky}+4\tilde{m}_5^2+\frac{\kappa^2 v_h^2}{3}m_5^2e^{2(2+\nu)ky}}}{8k},\nonumber\\
   &1&+ \frac{\sqrt{q^2e^{4(2+\nu)ky}+k^2+4\tilde{m}_{n}^2e^{2ky}+4\tilde{m}_5^2+\frac{\kappa^2 v_h^2}{3}m_5^2e^{2(2+\nu)ky}}}{4k},\nonumber\\
   &p&*\frac{e^{(2+\nu)ky}}{4k}\bigg]\bigg)
   \label{13S}
  \end{eqnarray}
  
  where we use the explicit form of warp factor ($A(y)$) and scalar field ($\xi(y)$). 
  However, from the continuity of $\tilde{\zeta}^{(n)}_{L,R}$ at $y=0$ and at $y=\pi r_c$, the KK mass spectrum in ST theory is obtained and given by,
   \begin{equation}
    m_n^2 = e^{-2A(\pi r_c)} [k^2(n^2 + 2n + 1) + m_5^2 e^{-\frac{\kappa}{\sqrt{3}}<\Phi>}]
    \label{mass 15S}
   \end{equation}
   
   where $n=1,2,3....$. The expression of KK mass tower is different in comparison to that of $F(R)$ theory. 
   However the difference is again small for $\kappa v_h < 1$.\\
   
   Using the solution of $\tilde{\zeta}^n_{L,R}(y)$ (in eqn. (\ref{13S})), we determine the coupling between 
   massive KK fermion modes and the radion field, given by
   \begin{eqnarray}
     &\tilde{\lambda}&^{(n)} = \sqrt{\frac{k}{24M^3}}e^{A(\pi r_c)} \exp{\bigg[-\frac{p}{4k}e^{(2+\nu)k\pi r_c}\bigg]}\nonumber\\
   &\exp&{\big[-\frac{1}{4k}\sqrt{q^2e^{4(2+\nu)k\pi r_c}+k^2+4\tilde{m}_{n}^2e^{2k\pi r_c}+4\tilde{m}_5^2
   +\frac{\kappa^2 v_h^2}{3}m_5^2e^{2(2+\nu)k\pi r_c}}\big]}\nonumber\\
   &\bigg[&HypergeometricU\bigg[\frac{5}{8} + \frac{qe^{2(2+\nu)k\pi r_c}}{8k}\nonumber\\
   &+&\frac{\sqrt{q^2e^{4(2+\nu)k\pi r_c}+k^2+4\tilde{m}_{n}^2e^{2k\pi r_c}+4\tilde{m}_5^2
   +\frac{\kappa^2 v_h^2}{3}m_5^2e^{2(2+\nu)k\pi r_c}}}{8k},\nonumber\\
   &1&+ \frac{\sqrt{q^2e^{4(2+\nu)k\pi r_c}+k^2+4\tilde{m}_{n}^2e^{2k\pi r_c}+4\tilde{m}_5^2+\frac{\kappa^2 v_h^2}{3}m_5^2e^{2(2+\nu)k\pi r_c}}}{4k},\nonumber\\
   &p&*\frac{e^{(2+\nu)k\pi r_c}}{4k}\bigg]\bigg]^2
  \label{16S}
   \end{eqnarray}
   
   where $\tilde{\lambda}^{(n)}$ is the coupling between $n$ th KK fermion mode and the radion field in St theory. 
   Similar to $F(R)$ theory, $\tilde{\lambda}^{(n)}$ decreases as $\alpha$ increases.\\
   
\begin{figure}[!h]
\begin{center}
 \centering
 \includegraphics[width=3.0in,height=2.0in]{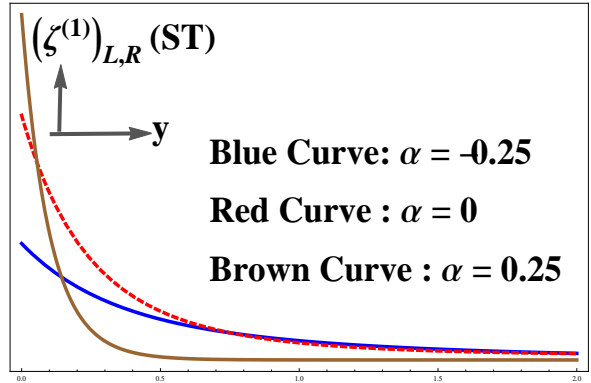}
 \caption{$\tilde{\zeta}^{(1)}_{L,R}(=\zeta^{(1)}_{L,R}(ST)$ in the diagram) vs $y$ for $M = 1$, $\Lambda = -1$ and $\frac{m_5}{k}=0.5$}
 \label{plot massive modeS}
\end{center}
\end{figure}

   Further from eqn.(\ref{13S}), one can find the dependence of localization 
   for massive KK fermion modes on the parameter $\alpha$. 
   From this, the behaviour of the first KK mode ($n=1$) is described in figure (\ref{plot massive modeS}).
   
Figure (\ref{plot massive modeS}) depicts that the wave function for 
first massive KK mode (in ST theory) gets more and more localized near Planck brane with increasing 
value of $\alpha$. As a result, the coupling parameter decreases near the visible 
brane as the value of $\alpha$ increases.\\  

At this stage, it deserves mentioning that although the solution of fermionic wave function as well as the expression 
of radion-fermion coupling become different (however the differences are indeed small for $\kappa v_h < 1$) 
in $F(R)$ and its corresponding ST theory, the localization property 
for both massless and massive KK fermionic mode remain identical in both the theories.

\section{Conclusion}
We consider a five dimensional AdS compactified warped geometry model with two 3-branes 
embedded within the spacetime. Due to large curvature ($\sim$Planck scale), the bulk spacetime 
is governed by higher curvature F(R) gravity. In the scenario of non constant curvature, we study in full generality, 
how the higher curvature terms affect 
the localization of a bulk fermion field within the entire spacetime where F(R) contains 
the next higher order curvature term to Einstein gravity i.e. $F(R)=R+\alpha R^2$. 
Moreover, we have also explored the localization property of the bulk fermion field in the conformally transformed 
scalar-tensor version of the model.  Our 
findings are as follows :

\begin{enumerate}
\item In $F(R)$ theory :
\begin{itemize}
 \item For massless KK mode:\\
 
  In the absence of bulk fermion mass, left and right chiral modes can be localized at different 
  regions in the spacetime by adjusting the value of higher curvature parameter $\alpha$. However, the localization 
  of both the chiral modes become sharper near TeV brane as the value of $\alpha$ increases.\\ 
  
  For non-zero bulk fermion mass, the left as well as right mode get more and more 
  localized as the higher curvature parameter becomes larger. Correspondingly the overlap of fermion 
  wave function with the visible brane increases with $\alpha$, which is depicted in figure (\ref{plot massless2}) 
  and figure (\ref{plot massless3}).\\
  
  The effective coupling between radion and zeroth 
  order fermionic KK mode is obtained (in eqn(\ref{coupling 4}) and eqn.(\ref{coupling 5})).  
  It is found that the radion-fermion coupling (for both left and right chiral mode) increases 
  with the increasing value of higher curvature parameter. This is a direct consequence of the fact 
  that the peak of the left and right chiral mode shift towards the visible brane 
  as the parameter $\alpha$ increases. To explore the radion phenomenology, the mass of radion field 
  is also determined in eqn.(\ref{radion_mass}). It is found that the contribution of radion in scattering amplitude 
  of fermions decreases as the value of the parameter $\alpha$ increases, which is depicted in figure (\ref{plot ratio}). Thus the 
  presence of higher curvature term reduces the possibility of radion detection in such scattering processes.\\
  
  Another important point to note is that for a fixed form of F(R), the left chiral mode has higher peak values on TeV brane 
  as the bulk fermion mass increases where as the localization of right chiral mode shows an opposite behaviour.  
  This is in agreement with \cite{tp_fermion,jm}.\\
 
 \item For massive KK mode :\\
 
  The requirement of solving the gauge hierarchy problem confines the mass 
  of higher KK modes at TeV scale. Moreover the mass squared gap ($\Delta m_n^2 =m_{n+1}^2-m_n^2$) depends linearly 
  on $n$ which is also evident from the mass spectrum in eqn. (\ref{mass spectrum}).\\
  
  The couplings between radion and massive KK fermionic mode are determined (in eqn. (\ref{coupling})) in the 
  presence of higher order curvature term (i.e. $F(R)=R+\alpha R^2$). 
  It is found that the coupling parameter decreases with the increasing higher curvature parameter.\\
  
  From the perspective of localization scenario, the wave function of massive KK 
  modes are localized near Planck brane which  
  increases with the order of KK mode. As a result, the couplings of the massive KK fermionic modes 
  with the visible brane matter fields become extremely weak and therefore drastically reduces the possibility 
  of finding the signatures of such massive fermion KK modes in TeV scale experiments.\\
\end{itemize}

\item In scalar-tensor theory :\\

Finally we investigate the localization property of the bulk fermion field in the scalar-tensor version of the theory. 
We observe that though the solution of fermionic wave function as well as the expression 
of radion-fermion coupling differ slightly from that in the corresponding $F(R)$ model, the localization property 
for both massless and massive KK fermionic mode remain identical in both the theories.
 
\end{enumerate}

\end{document}